\documentclass[twocolumn]{aastex62}
\usepackage{natbib}
\usepackage{graphicx}
\usepackage{amsmath}
\usepackage{txfonts}
\usepackage{amssymb}
\usepackage{multirow}

\received{\today}
\revised{TBD}
\accepted{TBD}

\submitjournal{ApJ}

\shorttitle{Host galaxy of IRAS 20181-2244}

\newcommand{\kms}{$\rm km\,s^{-1}$}
\newcommand{\ergs}{$\rm erg\,s^{-1}$}

\begin{document}

\title{\large The interacting late-type host galaxy of the radio-loud narrow-line Seyfert 1 IRAS 20181-2244}

\correspondingauthor{M. Berton}
\email{marco.berton@utu.fi}

\author[0000-0002-1058-9109]{M. Berton}
\affiliation{Finnish Centre for Astronomy with ESO (FINCA), University of Turku, Quantum, Vesilinnantie 5, 20014 University of Turku, Finland}
\affiliation{Aalto University Mets{\"a}hovi Radio Observatory, Mets{\"a}hovintie 114, FIN-02540 Kylm{\"a}l{\"a}, Finland}

\author[0000-0002-8549-4083]{E. Congiu}
\affiliation{Dipartimento di Fisica e Astronomia "G. Galilei", Universit\`a di Padova, Vicolo dell'Osservatorio 3, 35122 Padova, Italy}
\affiliation{INAF - Osservatorio Astronomico di Brera, via E. Bianchi 46, 23807 Merate (LC), Italy}
\affiliation{Las Campanas Observatory, Carnegie Institution of Washington, Colina El Pino Casilla 601, La Serena, Chile}

\author[0000-0001-9539-3940]{S. Ciroi}
\affiliation{Dipartimento di Fisica e Astronomia "G. Galilei", Universit\`a di Padova, Vicolo dell'Osservatorio 3, 35122 Padova, Italy}
\affiliation{INAF - Osservatorio Astronomico di Padova, Vicolo dell'Osservatorio 5, 35122 Padova, Italy}

\author[0000-0002-9214-4428]{S. Komossa}
\affiliation{Max-Planck-Institut f\"ur Radioastronomie, Auf dem H\"ugel 69, 53121 Bonn, Germany}

\author{M. Frezzato}
\affiliation{Dipartimento di Fisica e Astronomia "G. Galilei", Universit\`a di Padova, Vicolo dell'Osservatorio 3, 35122 Padova, Italy}

\author[0000-0003-0483-5083]{F. Di Mille}
\affiliation{Las Campanas Observatory, Carnegie Institution of Washington, Colina El Pino Casilla 601, La Serena, Chile}

\author[0000-0002-0658-644X]{S. Ant\'on}
\affiliation{CIDMA, Department of Physics, University of Aveiro, 3810-193, Aveiro, Portugal}

\author{R. Antonucci}
\affiliation{Department of Physics, University of California, Santa Barbara, CA 93106-9530, USA}

\author[0000-0002-2339-8264]{A. Caccianiga}
\affiliation{INAF - Osservatorio Astronomico di Brera, via E. Bianchi 46, 23807 Merate (LC), Italy}

\author{P. Coppi}
\affiliation{Yale Center for Astronomy and Astrophysics, Yale University, New Haven, CT 06520-8121, USA}

\author{E. J\"arvel\"a}
\affiliation{Aalto University Mets{\"a}hovi Radio Observatory, Mets{\"a}hovintie 114, FIN-02540 Kylm{\"a}l{\"a}, Finland}
\affiliation{Aalto University Department of Electronics and Nanoengineering, P.O. Box 15500, FI-00076 AALTO, Finland}

\author[0000-0003-0133-7644]{J. Kotilainen}
\affiliation{Finnish Centre for Astronomy with ESO (FINCA), University of Turku, Quantum, Vesilinnantie 5, 20014 University of Turku, Finland}

\author{A. L\"ahteenm\"aki}
\affiliation{Aalto University Mets{\"a}hovi Radio Observatory, Mets{\"a}hovintie 114, FIN-02540 Kylm{\"a}l{\"a}, Finland}
\affiliation{Aalto University Department of Electronics and Nanoengineering, P.O. Box 15500, FI-00076 AALTO, Finland}

\author{S. Mathur}
\affiliation{Department of Astronomy and Center for Cosmology and AstroParticle Physics, The Ohio State University, 140 West 18th Avenue, Columbus, OH 43210, USA}

\author{S. Chen}
\affiliation{Dipartimento di Fisica e Astronomia "G. Galilei", Universit\`a di Padova, Vicolo dell'Osservatorio 3, 35122 Padova, Italy}
\affiliation{Center for Astrophysics, Guangzhou University, 510006, Guangzhou, China}
\affiliation{Istituto Nazionale di Fisica Nucleare (INFN), Sezione di Padova, 35131, Padova, Italy}

\author{V. Cracco}
\affiliation{Dipartimento di Fisica e Astronomia "G. Galilei", Universit\`a di Padova, Vicolo dell'Osservatorio 3, 35122 Padova, Italy}

\author[0000-0001-8553-499X]{G. La Mura}
\affiliation{Laboratory of Instrumentation and Experimental Particle Physics, Av. Prof. Gama Pinto, 2 - 1649-003 Lisboa, Portugal}

\author[0000-0002-6717-1686]{P. Rafanelli}
\affiliation{Dipartimento di Fisica e Astronomia "G. Galilei", Universit\`a di Padova, Vicolo dell'Osservatorio 3, 35122 Padova, Italy}

\begin{abstract}
Narrow-line Seyfert 1 galaxies (NLS1s) are a class of active galactic nuclei (AGN) which are known to be one of the few sources of $\gamma$ rays, which originate in a relativistic beamed jet. Becuase of their relatively large distance, a poorly investigated aspect of these jetted NLS1s is their environment, and in particular their host galaxy. In this work we present the results of a morphological analysis of the host galaxy of the jetted NLS1 IRAS 20181-2244 observed with the 6.5m Baade Telescope of the Las Campanas Observatory. The GALFIT analysis ran on the Ks image, along with additional spectroscopic observations performed with the Nordic Optical Telescope, clearly revealed the presence of an interacting system of two galaxies. The data suggest that this NLS1 is hosted by a late-type galaxy, although the result is not conclusive. This analysis, along with other results in the literature, might suggest that two populations of jetted NLS1 exist. Further morphological studies are needed to confirm or disprove this hypothesis.
\end{abstract}


\section{Introduction}

For many years it was believed that radio-loud\footnote{Radio-loudness is defined as the ratio $R$ between the $5$ GHz and the optical B-band flux densities \citep{Kellermann89}. A source is considered radio-loud if $R>10$. Otherwise, it is radio-quiet. The significance of this parameter is highly debated (e.g., see \citealp{Padovani17, Lahteenmaki18}).} active galactic nuclei (AGN), and quasars in particular, were hosted in giant elliptical galaxies \citep{Laor00, Chiaberge11}. 
Powerful relativistic jets were indeed observed in systems hosting very massive black holes that, as shown by the M$_{BH}$-$\sigma_*$ relation \citep{Ferrarese00}, are found in ellipticals.
The jet power, however, directly scales with the black hole mass \citep{Heinz03,Foschini11b}. 
Relativistic jets launched by high-mass black holes are more powerful and easier to detect.
More recent studies, in fact, detected relativistic jets in spiral galaxies \citep[e.g.,][]{Mao15}. 
Furthermore, the high sensitivity of the \textit{Fermi/LAT} Satellite led to the discovery of $\gamma$-ray emission coming from narrow-line Seyfert 1 galaxies (NLS1, \citealp{Abdo09a,Abdo09c}), indicating the presence of relativistic jets in a fraction of these AGN. \par 
Classified according to their spectral properties \citep{Osterbrock87, Goodrich89}, NLS1s show narrow permitted lines which, unlike in type 2 AGN, are not attributed to obscuration. The presence of strong Fe II multiplets in the spectrum, in fact, indicates that the broad-line region (BLR) is directly visible. NLS1s are often considered young (i.e., galactic nuclei in their first activity phase, \cite{Mathur00, Mathur01, Kawakatu07, Wang07, Foschini15, Berton17, Berton18a}) or rejuvenated AGN \citep{Mathur12}. These objects harbor a relatively low-mass black hole ($10^6$-$10^8$ M$_\odot$ estimated via single epoch virial technique, \citealp{Peterson11}) accreting close to the Eddington limit \citep{Boroson92} growing fast and evolving toward high-mass objects \citep{Mathur01}. This relatively low gravitational potential induces a low rotational velocity in the gas, which translates into the narrow lines. 
However, a disk-like geometry of the BLR might also account for the narrow lines if observed pole-on, because of the lack of Doppler broadening. 
If in some sources inclination plays a major role, the virial technique may underestimate their black hole mass.
These low-inclination NLS1s then may not harbor a low-mass black hole and, possibly, be different with respect to other NLS1s \citep{Sulentic00, Decarli08, Shen14} . \par
A way to distinguish between low-mass and low-inclination NLS1s is by means of host galaxy studies. 
The host galaxy can be an independent indicator of the black hole mass. 
While late-type hosts can be associated with lower masses, in fact, ellipticals typically harbor high-mass black holes \citep[e.g.,][and references therein]{Salucci00, Kormendy13}.
For this reason, an NLS1 with an elliptical host galaxy may have a high-mass black hole, and possibly a different nature.\par
A number of very nearby radio-quiet (or non-jetted, \citealp{Padovani17}) NLS1s are hosted by spiral galaxies \citep{Deo06, Mathur12}, with a strong ongoing star formation \citep{Sani10} and a pseudobulge due to secular evolution \citep{Orbandexivry11, Mathur12}. 
In radio-loud (or jetted, namely harboring a relativistic jet) NLS1s, instead, the host galaxy nature is still under debate, with controversial results supporting both elliptical and disk hosts \citep{Anton08, Leontavares14, Kotilainen16, Olguiniglesias17, Dammando17, Dammando18, Jarvela18}. \par
In this paper we present the results of a study carried out on the radio-loud NLS1s \citep{Komossa06} IRAS 20181-2244 (z$=$0.185) using the Magellan 6.5m Walter Baade telescope of the Las Campanas Observatory, and we will put them in the framework of the previously described scenario. 
In Sect.\,\ref{sec:object} we describe the source, in Sect.\,\ref{sec:datared} we describe the data reduction and analysis, while in Sect.\,\ref{sec:discussion} we finally discuss the results. 
Throughout this work, we adopt a standard $\Lambda$CDM cosmology, with a Hubble constant $H_0 = 70$ \kms Mpc$^{-1}$, and $\Omega_\Lambda = 0.73$ \citep{Komatsu11}.

\begin{figure}[t!]
\centering
\includegraphics[width=\columnwidth]{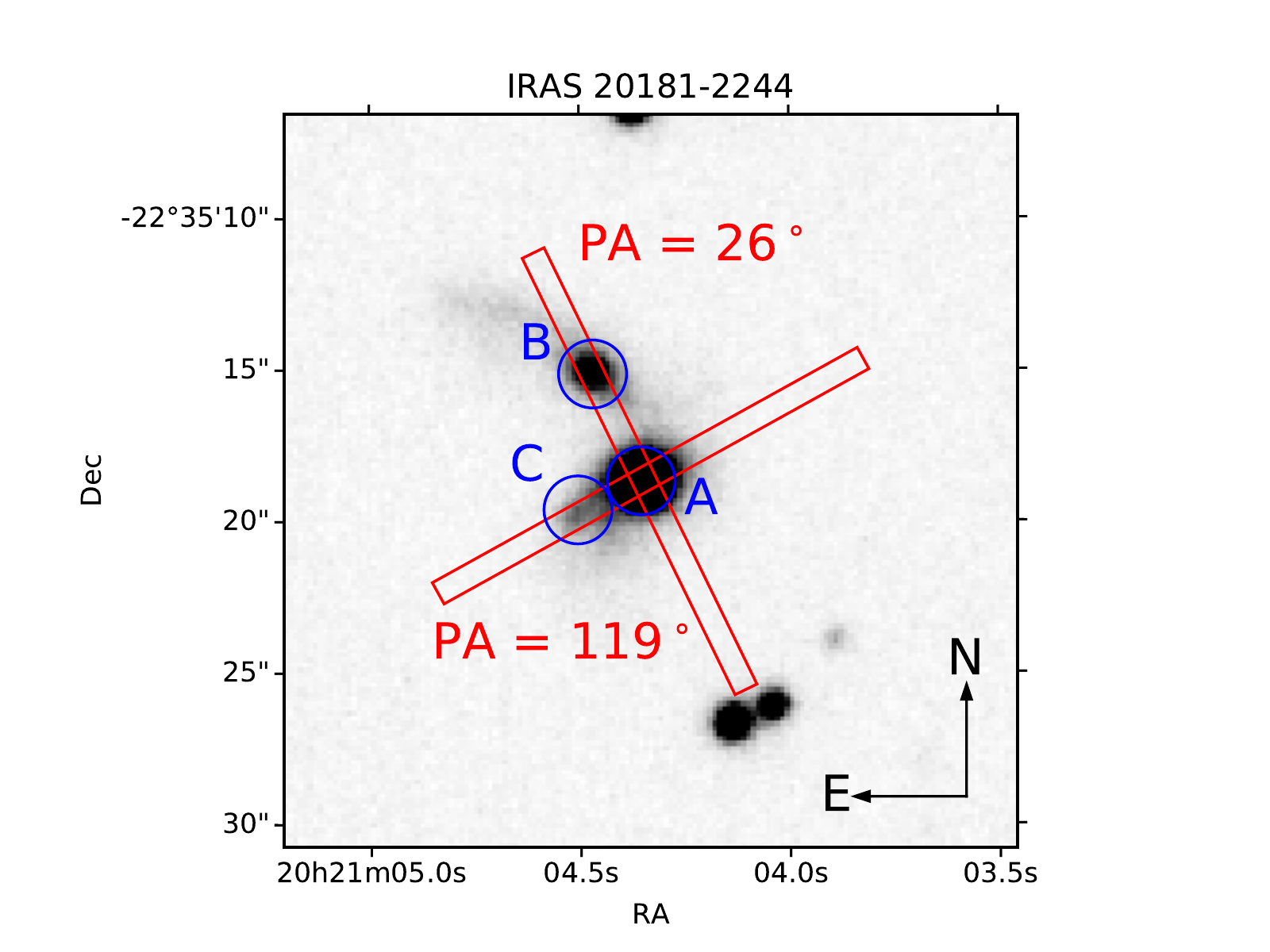} 
\caption{Finding chart of IRAS 20181-2244 (marked as A) and its putative companion sources B and C. The red boxes represent the position of the slit during the NOT observations. The two sources on the south west are stars.}
\label{fig:finding_chart}
\end{figure}

\section{IRAS 20181-2244}
\label{sec:object}
Initially classified as a Seyfert 2 \citep{Elizalde94}, it was reclassified as a NLS1 by \citet{Halpern98}.
Its optical B-band magnitude derived from the USNO-B catalog \citep{Monet03} is m$_{\rm B} =$ 16.86, which provides an absolute magnitude of M$_{\rm B} = -$22.81.
Hence, IRAS 20181-2244 could formally be classified as a narrow-line type 1 quasar \citep{Schmidt83}. 
However, to avoid confusion with similar sources in the literature, in the following we will stick to the NLS1 classification. 
For many years it has been known as a bright X-ray source from early ROSAT observations \citep{Boller92}, as many other NLS1s \citep{Boller96}, but it was recognized as a bright radio source only later \citep{Komossa06}, with a luminosity of 3.6$\times$10$^{40}$ \ergs\ at 1.4 GHz. 
According to \citet{Komossa06} and \citet{Foschini15}, this object has a radio spectral index of 0.50$\pm$0.07 (F$_\nu \propto \nu^{-\alpha}$) around 1.4 GHz.
Therefore it lies at the divide between flat- and steep-spectrum radio-loud NLS1s, with an estimated radio-loudness of $\sim$45 \citep{Chen18}. 
Its black hole mass estimates span between 3$\times$10$^6$ M$_\odot$ to 3.75$\times$10$^7$ M$_\odot$ using single epoch spectroscopy \citep{Komossa06} and optical magnitudes \citep{Foschini15}, respectively. 
Both authors provide an Eddington ratio estimate of 0.6, due to a different calculation of the bolometric luminosity. 
Both mass and Eddington ratio are in the typical range for NLS1s. \par
One of the most interesting characteristics of IRAS 20181-2244 is its remarkably high star formation rate (SFR), up to $\sim$300 M$_\odot$ yr$^{-1}$ \citep{Caccianiga15}. 
This value is the highest among known radio-loud NLS1s, and typical for sources in the ultraluminous infrared galaxies (ULIRG) regime \citep{Sanders88}. 
Despite this surprisingly high value, the SFR is not enough to explain the totality of the radio emission. 
Indeed, assuming a SFR of 300 M$_\sun$ yr$^{-1}$ and using the relation between SFR and radio luminosity at 1.4GHz of \citet{Condon02}, we derive an expected radio luminosity due to SF that can explain at most $\sim$30\% of the observed luminosity. Therefore, the presence of the non-thermal emission of a relativistic jet is required. We note that even considering the possible amount of radio emission due to SF, the radio-loudness parameter would be lower but still above the formal limit of radio-loud AGN.
This makes IRAS 20181-2244 a unique source worth of a detailed investigation.

\section{Data reduction and analysis}
\label{sec:datared}

We observed IRAS 20181-2244 on 2016-10-13 with the FourStar instrument of the Walter Baade 6.5m telescope of the Las Campanas Observatory. 
We acquired images in J, H, and Ks bands, with a total exposure time of 256 seconds in Ks, 640 seconds in H, and 640 seconds in J band. 
The seeing was 0.67$^{\prime\prime}$ in the Ks band and 0.72$^{\prime\prime}$ in J and H bands. 
The detector scale is 0.159$^{\prime\prime}$ px$^{-1}$, corresponding to 0.473 kpc px$^{-1}$.
We performed a standard reduction using \texttt{IRAF}\footnote{http://iraf.noao.edu/}, with bias and flat-field correction, followed by the alignment, sky subtraction, fringing removal, and combination of the images in each filter. 
We later estimated the zeropoint (zp) for photometric calibration of each filter by comparing the instrumental magnitudes of some of the stars in the field, measured with PSF photometry, with their 2MASS magnitudes.
The resulting zeropoints were zp$_H = 25.64 \pm 0.29$, zp$_J = 26.01 \pm 0.14$, and zp$_{Ks} = 25.12 \pm 0.18$.\par
The source has a rather complicated structure. 
In addition to the AGN and its host galaxy (component A in Fig.~\ref{fig:finding_chart}), we observe a nearby galaxy at position angle (P.A.) 26$^\circ$ (component B in Fig.~\ref{fig:finding_chart}). 
We also identify an extended emission at P.A. 119$^\circ$ which might be a point-like source distorted by seeing, and therefore another possible galaxy (component C in Fig.~\ref{fig:finding_chart}). 
All the other objects surrounding the system are instead confirmed stars.
To investigate the nature of the putative companions, we decided to carry out spectroscopic observations to determine their redshifts. 

\subsection{Spectroscopy}
\begin{figure}[t!]
\centering
\includegraphics[width=\columnwidth]{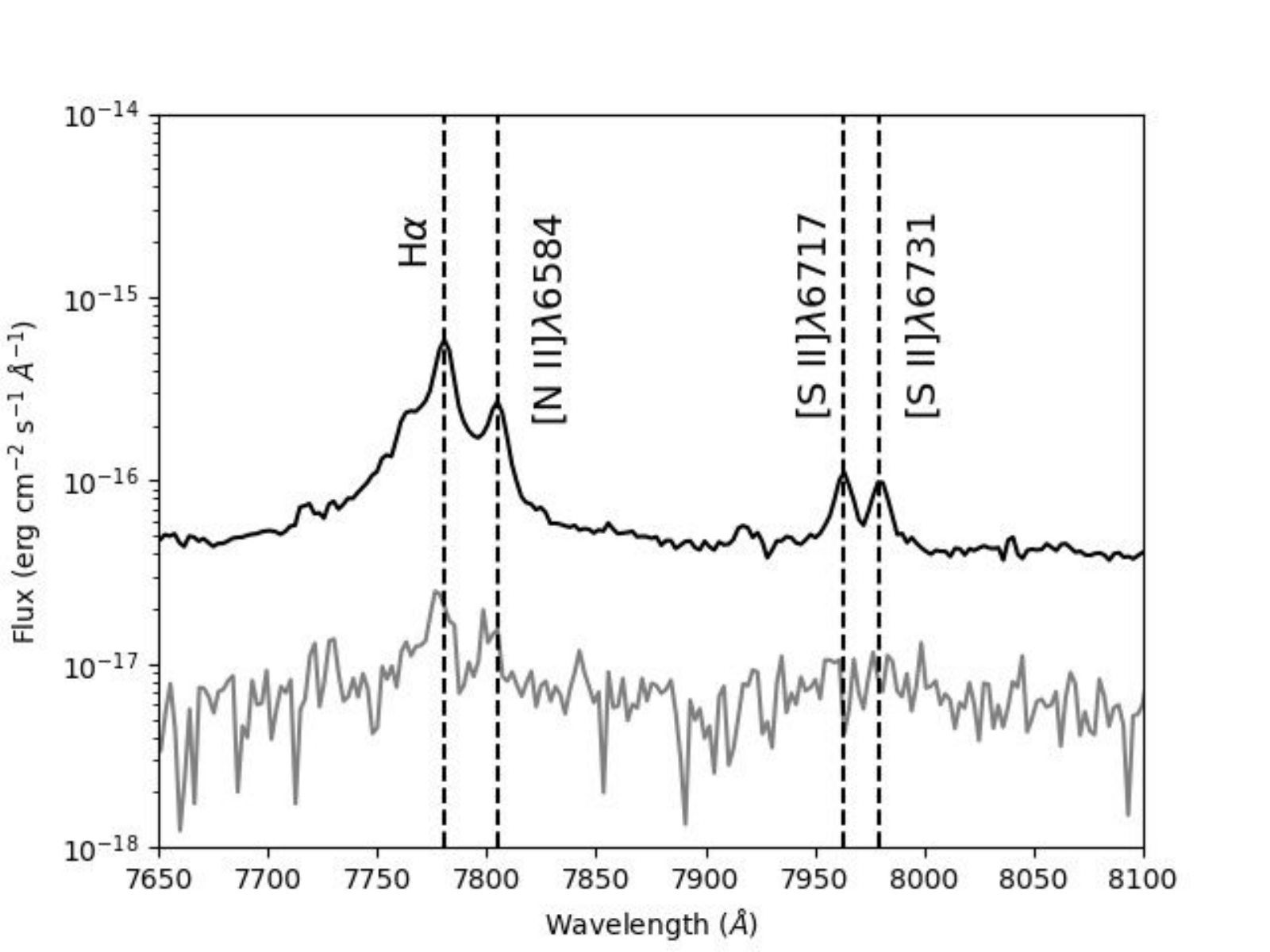} 
\caption{Spectra of IRAS 20181-2244 (black line) and its putative companion B (grey line) at P.A. $= 26^\circ$ in the H$\alpha$ region. The flux scale is logarithmic to enhance the lines of the companion. The dashed vertical lines mark the most prominent lines.}
\label{fig:spectrum_companion}
\end{figure}
\begin{figure*}[ht!]
\centering
\includegraphics[width=0.8\textwidth]{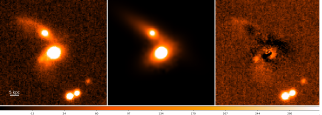} 
\includegraphics[width=0.8\textwidth]{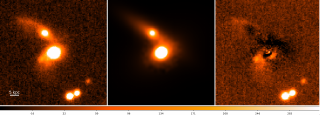} 
\caption{\textbf{Left:} Ks band image of IRAS 20181-2244; \textbf{middle:} GALFIT model of the NLS1 and its companion; \textbf{right:} residuals of the GALFIT modeling. The top row refers to the disk host model, the bottom row to the elliptical host model. The color scale is shown below both rows. Spatial scale at the AGN redshift is indicated in both images. Orientation as in Fig.~\ref{fig:finding_chart}. }
\label{fig:source}
\end{figure*}
We observed IRAS 20181-2244 with the Nordic Optical Telescope (NOT) on 2017-08-04 and 2017-08-11.
To this aim we acquired two spectra at two different position angles (P.A. $= 26^\circ$ and P.A. $= 119^\circ$) to include in each observation the nucleus of the AGN and one of the two putative companion sources.
The observations were performed using the Andalucia Faint Object Spectrograph and Camera (ALFOSC).
We used a 0.5$^{\prime\prime}$ slit and the grism 20, covering a spectral range from 5650 to 10150 \AA\ with a resolution R $\sim$ 1540.
For each P.A. we acquired 3 spectra for a total exposure time of 1800 s at P.A. $= 26^\circ$ and 2700 s at P.A. $= 119^\circ$.
A spectrum of the standard star Feige 110 was acquired for flux calibration while a ThAr lamp and standard flat fields were acquired for wavelength calibration and flat field correction.
A standard calibration procedure has been carried out using IRAF tasks.\par
In the NLS1 spectrum all the most prominent Balmer lines, along with [O III] $\lambda\lambda$4959,5007, He I $\lambda$5876, [O I] $\lambda$6300, [N II] $\lambda\lambda$6548,6584, [S II] $\lambda\lambda$6716,6731 are detected.
After correcting for telluric absorption, to measure the width of the broad line we decomposed the H$\alpha$ line profile using three Gaussians.
Two Gaussians are needed to represent the broad component, because its complex profile is suggestive of two different kinematic regions, as usually observed in broad-line profiles \citep[e.g., ][]{Popovic04}. 
A third one is instead used to represent the narrow component. 
We obtained FWHM$_{\rm broad} \sim$ 1500 km s$^{-1}$, thus confirming the NLS1 nature of this source.\par
At P.A. $= 26^\circ$ the AGN emission lines seem to be extended up to a region where we clearly detect a faint continuum and weak emission lines (H$\alpha$ and [N II]$\lambda$6584 are detected, see Fig.~\ref{fig:spectrum_companion}). 
Both lines and continuum are located $\sim 3.5^{\prime\prime}$ away from the NLS1 nucleus, in the same position as source B in Fig.~\ref{fig:finding_chart}. 
We tested the hypothesis that these lines originate in the AGN and are extended over source B because of the seeing. 
However, neither lines nor continuum were detected on the opposite side of source A, thus ruling out the seeing origin. 
Another possibility is that the observed lines are due to the extended narrow-line region (ENLR) of the AGN. 
Nevertheless, no emission lines shifted with respect to those of the putative ENLR were detected in the spectrum. 
Therefore, if this is the case and source B emits some lines (a likely possibility given its disk nature), they are blended with those of the ENLR because of the vicinity of the two sources. 
The final option is simply that the observed lines originate in source B, confirming that both A and B are at the same z = 0.185. \par
In the spectrum at P.A. $= 119^\circ$ some emission-lines from IRAS 20181-2244 are instead projected onto source C.
This source, in fact, is much closer to the AGN with respect to the previous case, and the observations were performed with a seeing around $1.5^{\prime\prime}$. 
However, no emission or absorption lines are detected from source C itself, implying that it is either an inactive galaxy, or that it has faint emission lines at the same redshift as the NLS1 which are not detectable. 
In conclusion, nothing can be said about the nature of source C with our data.

\subsection{Imaging}
\label{sec:imaging}
\begin{table}[t!]
\caption{Parameters derived from GALFIT fitting. Each component is indicated in boldface. Columns: (1) right ascension (hh:mm:ss); (2) declination (dd:mm:ss); (3) integrated Ks magnitude; (4) effective radius in kpc (and arcsec); (5) S\'ersic index.}
\label{tab:parameters}
\centering
\begin{tabular}{l c c c c c}
\hline\hline
R.A. & Dec. & mag. & R$_e$ & n \\
\hline
\multicolumn{2}{l}{\bf{PSF A (disk case)}} \\
20:21:04.4 & -22:35:18.7 & 13.41 &  {} & {} \\
\hline
\multicolumn{2}{l}{\bf{S\'ersic A (disk case)}} \\
\multirow{2}{1.5cm} {20:21:04.4} & \multirow{2}{1.5cm} {-22:35:19.0} & \multirow{2}{0.7cm} {14.22} & 5.56 & 1 \\
 & & & (1.87) & & \\
\hline
\multicolumn{2}{l}{\bf{PSF A (elliptical case)}} \\
20:21:04.4 & -22:35:18.7 & 13.73 &  {} & {} \\
\hline
\multicolumn{2}{l}{\bf{S\'ersic A (elliptical case)}} \\
\multirow{2}{1.5cm} {20:21:04.4} & \multirow{2}{1.5cm} {-22:35:18.7} & \multirow{2}{0.7cm} {13.71} & 7.44 & 4  \\
 & & & (2.50) & & \\
\hline
\multicolumn{2}{l}{\bf{S\'ersic B (bulge)}} \\
\multirow{2}{1.5cm} {20:21:04.5} & \multirow{2}{1.5cm}{-22:35:15.0} & \multirow{2}{0.7cm} {16.24} & 0.29 & \multirow{2}{0.5cm} {0.51} \\
& & & (0.10) & \\
\hline
\multicolumn{2}{l}{\bf{S\'ersic B (host)}} \\
\multirow{2}{1.5cm} {20:21:04.5} &  \multirow{2}{1.5cm}{-22:35:15.0} & \multirow{2}{0.7cm} {14.82} & 5.81 & 1  \\
 & & & (1.95) & \\
\hline\hline
\end{tabular}
\end{table}

\begin{figure*}[t!]
\centering
\includegraphics[width=\columnwidth]{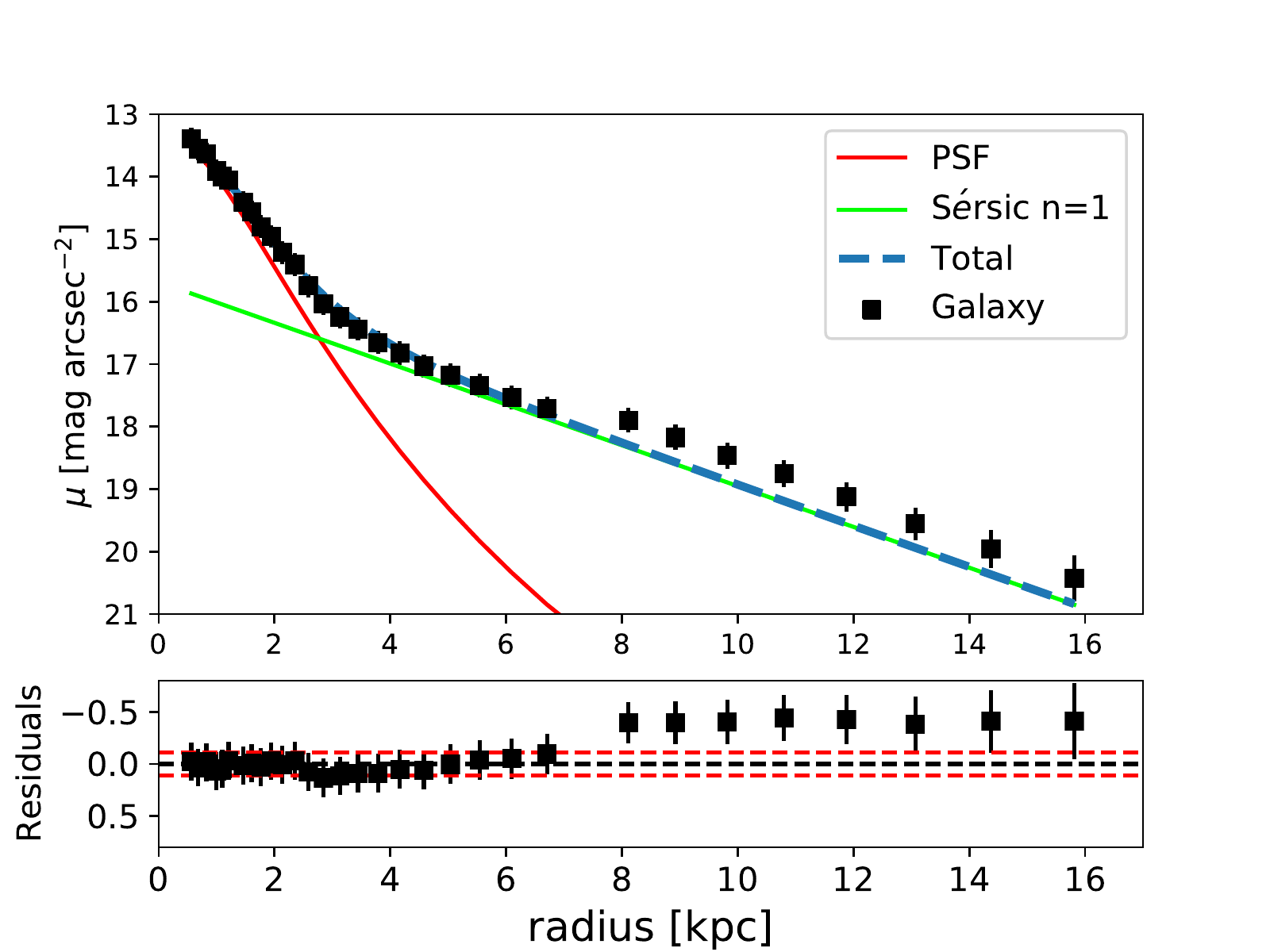} 
\includegraphics[width=\columnwidth]{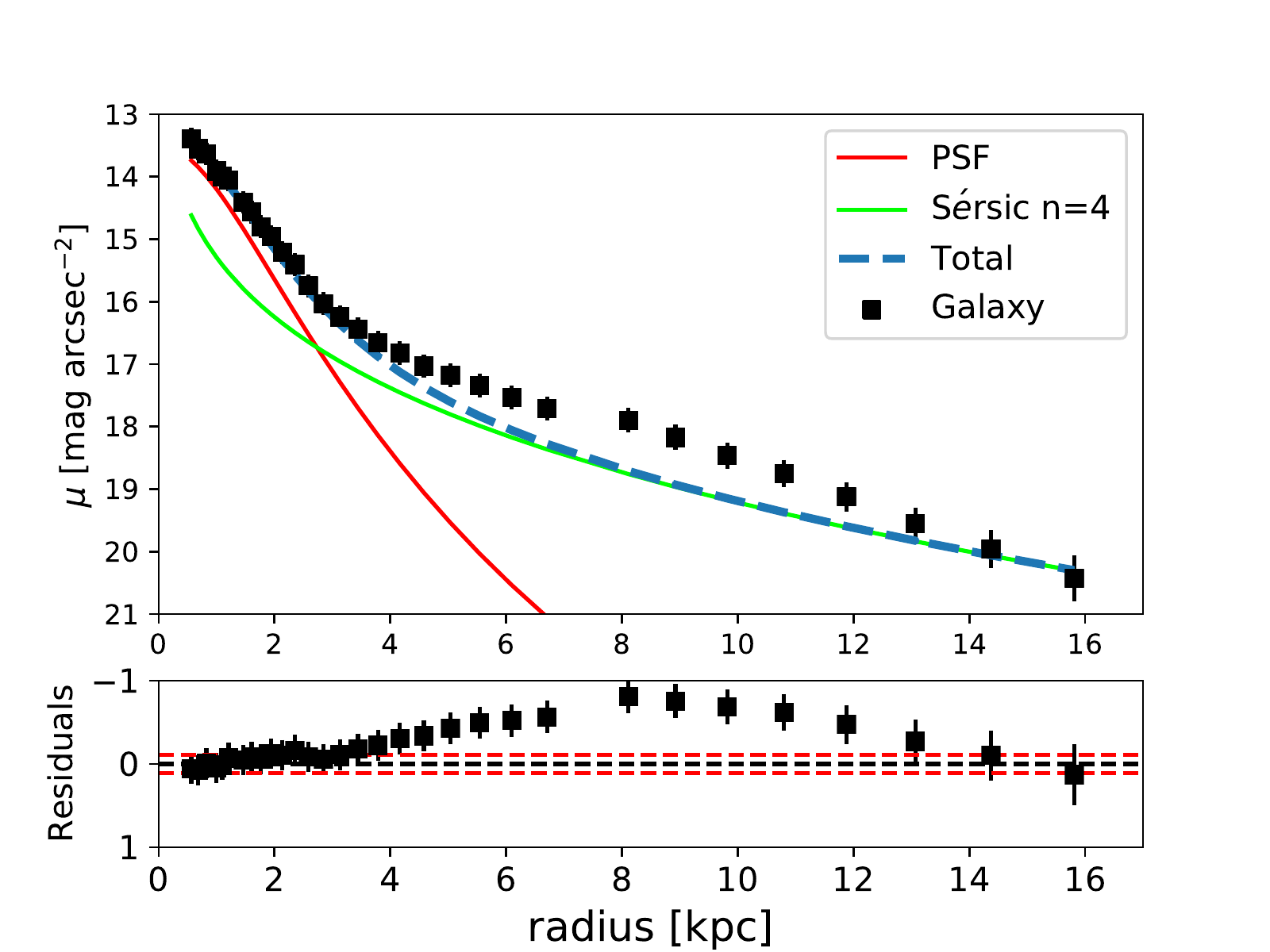} 
\caption{Brightness profile of IRAS 20181-2244 modeling. The input parameters are those derived by GALFIT. \textbf{Left panel:} on top, the black squared points represent the isophotes of the galaxy, the red solid line represents the PSF, the green solid line represents the host galaxy model with a S\'ersic index n $=$1 (exponential disk), and the blue dashed line indicates the total model. Bottom, residuals after model subraction; the dashed black line represents the zero level, the dashed red lines indicate the background noise level. \textbf{Right panel:} as in the left panel, but the host galaxy model has a S\'ersic index n $=$4. }
\label{fig:brightness_profile}
\end{figure*}

To investigate the morphology of our target we used GALFIT \citep{Peng02, Peng10}. 
We carried out the procedure only in the Ks band.
Given the worse seeing conditions during the observations (0.75$^{\prime\prime}$ in H, J band vs 0.67$^{\prime\prime}$ in Ks) and the lower quality of the H and J band due to strong fringing, we could not obtain any additional information in the other filters with respect to the Ks band. 
We initially reproduced the AGN with a PSF described by a Moffat function to find its exact position. 
The PSF profile was estimated by measuring and averaging the PSFs of 20 non-saturated stars in the field-of-view. 
As a second step we reproduced source B, fitting it with a combination of two S\'ersic profiles, one for the bulge and one for the exponential disk. 
After obtaining a suitable result, we added a S\'ersic profile to reproduce the host galaxy of the AGN. 
Finally, we also tried to model component C using a S\'ersic profile, but we could not obtain any reliable result. 
The parameters derived from the fitting procedure are summarized in Table~\ref{tab:parameters}.
Given that the nuclear PSF is dominating over the center of the galaxy because of the seeing, we decided to avoid trying to reproduce both the bulge and the host, but to focus only on the latter.

\begin{figure}[t!]
\centering
\includegraphics[width=\columnwidth]{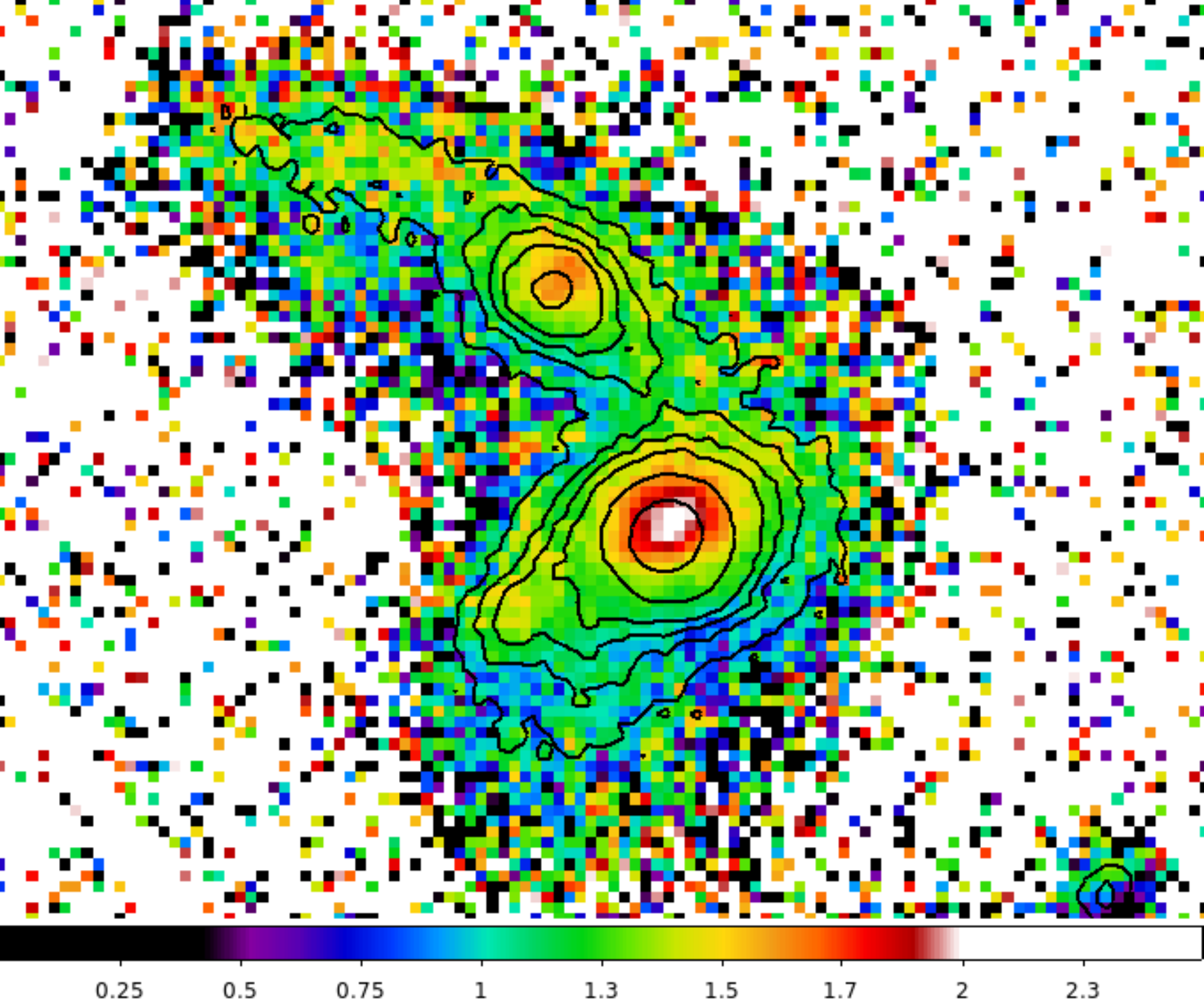} 
\caption{J$-$Ks color of IRAS 20181-2244. The color scale is reported at the bottom of the image. The contour levels represent 10, 20, 30, 45, 100, 400 times the rms calculated on the Ks band image. }
\label{fig:J-K}
\end{figure}
With no constraint, its S\'ersic index converged to 0.29, suggesting a late-type nature for the host. 
To derive properly the physical parameters of the host, we re-fitted it with an exponential disk (S\'ersic index n = 1), obtaining a reduced-$\chi^2$ ($\chi^2_\nu) = $ 1.567. 
The original Ks image, the model produced by GALFIT, and the residual image are shown in the top row of Fig.~\ref{fig:source}. 
For further confirmation of GALFIT results, we also extracted the brightness profile of the AGN host galaxy using the IRAF task \texttt{ellipse}, while masking companion B.  
The result is shown in the left panel of Fig.~\ref{fig:brightness_profile}. 
The parameters used to reproduce the profile are those already derived by GALFIT. \par
We also tried to reproduce the host galaxy profile with a S\'ersic with index n = 4, which represents the brigthness profile of an elliptical galaxy, along with the nuclear PSF.
This attempt led to $\chi^2_\nu=$1.739, higher than that obtained for the disk model. 
The GALFIT modeling is shown in the bottom row of Fig.~\ref{fig:source}, while the brightness profile is displayed in the right panel of Fig.~\ref{fig:brightness_profile}. \par
We also estimated the J$-$Ks color of the galaxy. 
Given that the shape of the near infrared spectrum is not known, we did not K-correct the different bands. 
The resulting map is shown in Fig.~\ref{fig:J-K}.

\section{Discussion}
\label{sec:discussion}

Because of the seeing, the nuclear region is dominated by the PSF of the AGN. 
This prevented any modeling of a potential bulge component.
As shown in Fig.~\ref{fig:brightness_profile}, in fact, no additional component is needed beside the host and the PSF to reproduce the brightness profile of the inner regions of this galaxy. \par
Regarding the host, as shown in Table~\ref{tab:parameters} and Fig.~\ref{fig:source} and \ref{fig:brightness_profile}, the PSF + exponential disk model can reproduce quite well the outer regions of the galaxy up to 8 kpc, but it fails at larger radii, where a significant deviation is observed. 
However, this flux excess may be due to the presence of component C, which was not modeled and lies at $\sim$8 kpc ($\sim$2.6$^{\prime\prime}$) away from the nucleus (assuming it is at the same redshift as the AGN). \par
The elliptical host + PSF model instead seems to reproduce the data only up to 4 kpc from the nucleus. 
At larger distances, the data lie systematically above the model predictions, and only at very large radii ($>$13 kpc) the model seems to be again in agreement with the observations. 
Another weakness of this model is that the brightness profile of the elliptical host is significantly higher in the nuclear region with respect to the disk, and its peak magnitude is almost comparable to that of the AGN. 
If true, we would expect to observe in the optical spectrum at least some absorption lines and the 4000\AA\ break that, instead, are not observed. 
Therefore, despite some data limitations, we are inclined to believe that the disk model is a better representation of the host galaxy.  \par
As mentioned above, the significant deviation from the exponential profile at large radii may be associated with component C. 
Its nature is not certain yet. 
The optical spectrum of this region, given its relatively small distance from the nucleus, was dominated by the AGN, therefore it was not possible to fully determine whether it is a bright star forming region, or a small galaxy interacting with the AGN host, or a tidal tail produced by the interaction between components A and B. 
Conversely, source B is a late-type galaxy, well reproduced by a S\'ersic profile for the bulge with the addition of an exponential profile for the disk. 
Also this object shows a luminosity excess in the residuals, which is suggestive of a tidal tail produced by its interaction with the AGN host. \par
The color index J$-$Ks provides some additional information. 
In the NLS1 nucleus its value is $\sim$2.1. 
This result is not far from typical colors observed in other AGN \citep{Fischer06, Masci10}, even if this one seems to be slighly redder when compared to other quasars \citep{Leipski07}. 
A small fraction of this very red color (around $\sim$0.2) could be attributed to the lack of K-correction in our data, but despite this the nucleus is definitely very red. 
The reason for this could be a combination of dust absorption and host dilution. 
The nucleus of the companion galaxy is also fairly red ($\sim$1.6), showing that dust might be very abundant in both of these interacting galaxies. 
The colors could also be an indication that there is a significant ongoing star formation activity in the reddest parts of the system. 
The star formation indeed is responsible for a significant Pa$\alpha$ emission which, because of the redshift, falls in the Ks band lowering its magnitude. 
In particular, the inner 3 kpc of the nucleus, source C, and the tidal tail of source B might all possibly contribute to the extreme star formation rate observed in this galaxy by \citet{Caccianiga15}. \par
The ongoing interaction which we observe might play a key role in the relativistic jet production. 
In high-mass jetted AGN, for example, merging and interaction are frequently observed \citep{Chiaberge15}.  
If jetted AGN are associated with rapidly spinning black holes, merging can help to spin up the black hole via several accretion episodes, thus triggering the relativistic jet launching. 
If this is true, among NLS1s, non-jetted sources might have a lower spin with respect to jetted NLS1s \citep[e.g.,][and references therein]{Chiaberge17}, since they tend not to show merging episodes \citep{Krongold01, Xu12}. 
This lack of frequent interactions found among non-jetted NLS1s might be a consequence of their large-scale enviroment, which is different with respect to that of jetted NLS1s \citep{Jarvela17a, Jarvela17}. 
Galaxies which reside in a dense environment, on average, have a higher probability to interact with other sources than galaxies located in voids. Even if the relative velocity between galaxies and the position of each galaxy within its environment also play a role in the process, we can conclude that a dense environment may usually produce a higher fraction of jetted sources. \par
In conclusion, if the disk host hypothesis for IRAS 20181-2244 is correct, this NLS1 (we remark that formally this object is classifiable as a quasar) is hosted by a galaxy with a scale radius of 3.3 kpc, slightly larger than that of the Milky Way in the same band \citep{Porcel98}. 
This source, then, may belong to the class of disk-hosted jetted NLS1s, as some other similar objects found in the literature \citep{Anton08, Kotilainen16, Olguiniglesias17, Jarvela18}. 
The presence of jetted NLS1s hosted also in elliptical galaxies \citep{Dammando17, Dammando18} suggests that both classes of jetted NLS1s exist. 
In a similar fashion, evidence for two different classes of NLS1s has already been discussed in the literature, based for instance on the spread of X-ray slopes, on the range of Eddington ratios, and the strength (or weakness) of Fe II emission \citep[e.g.,][and references therein]{Xu99, Sulentic00, Williams04, Mathur05a, Mathur05b}. 
We speculate that, while disk-hosted sources are characterized by young age and fast evolution \citep{Foschini15, Berton16c, Berton17, Komossa18}, the others may be instead low-inclination, and possibly not young, sources that "mimic" the behavior of genuine NLS1s. 
Further morphological studies on jetted and non-jetted NLS1s are anyway needed to confirm or disprove this scenario, and to shed more light on the formation of relativistic jets in this intriguing class of AGN.

\begin{acknowledgements}
\footnotesize
\textbf{Acknowledgements.} This paper is based on observations made with the Nordic Optical Telescope, operated by the Nordic Optical Telescope Scientific Association at the Observatorio del Roque de los Muchachos, La Palma, Spain, of the Instituto de Astrofisica de Canarias.
JK acknowledges financial support from the Academy of Finland, grant 311438. 
SA acknowledges support from CIDMA strategic project (UID/MAT/04106/2013), ENGAGE SKA (POCI-01-0145-FEDER-022217), funded by COMPETE 2020 and FCT, Portugal.
We are grateful to M. Baratella, I. Pagotto, and E. Sissa for helpful suggestions. 
This research has made use of the NASA/IP.A.C Extragalactic Database (NED) which is operated by the Jet Propulsion Laboratory, California Institute of Technology, under contract with the National Aeronautics and  Space Administration.
The National Radio Astronomy Observatory is a facility of the National Science Foundation operated under cooperative agreement by Associated Universities, Inc. 
Funding for the Sloan Digital Sky Survey has been provided by the Alfred P. Sloan Foundation, and the U.S. Department of Energy Office of Science. The SDSS web site is \texttt{http://www.sdss.org}. 
SDSS-III is managed by the Astrophysical Research Consortium for the Participating Institutions of the SDSS-III Collaboration including the University of Arizona, the Brazilian Participation Group, Brookhaven National Laboratory, Carnegie Mellon University, University of Florida, the French Participation Group, the German Participation Group, Harvard University, the Instituto de Astrofisica de Canarias, the Michigan State/Notre Dame/JINA Participation Group, Johns Hopkins University, Lawrence Berkeley National Laboratory, Max Planck Institute for Astrophysics, Max Planck Institute for Extraterrestrial Physics, New Mexico State University, University of Portsmouth, Princeton University, the Spanish Participation Group, University of Tokyo, University of Utah, Vanderbilt University, University of Virginia, University of Washington, and Yale University. 
This research has made use of "Aladin sky atlas" developed at CDS, Strasbourg Observatory, France.
\end{acknowledgements}

\bibliography{../biblio}

\begin{thebibliography}{}
\expandafter\ifx\csname natexlab\endcsname\relax\def\natexlab#1{#1}\fi
\providecommand{\url}[1]{\href{#1}{#1}}
\providecommand{\dodoi}[1]{doi:~\href{http://doi.org/#1}{\nolinkurl{#1}}}
\providecommand{\doeprint}[1]{\href{http://ascl.net/#1}{\nolinkurl{http://ascl.net/#1}}}
\providecommand{\doarXiv}[1]{\href{https://arxiv.org/abs/#1}{\nolinkurl{https://arxiv.org/abs/#1}}}

\bibitem[{{Abdo} {et~al.}(2009{\natexlab{a}}){Abdo}, {Ackermann}, {Ajello},
  {Axelsson}, {Baldini}, {Ballet}, {Barbiellini}, {Bastieri}, {Battelino}, \&
  {Baughman}}]{Abdo09a}
{Abdo}, A.~A., {Ackermann}, M., {Ajello}, M., {et~al.} 2009{\natexlab{a}},
  \apj, 699, 976, \dodoi{10.1088/0004-637X/699/2/976}

\bibitem[{{Abdo} {et~al.}(2009{\natexlab{b}}){Abdo}, {Ackermann}, {Ajello},
  {Baldini}, {Ballet}, {Barbiellini}, {Bastieri}, {Bechtol}, {Bellazzini}, \&
  {Berenji}}]{Abdo09c}
---. 2009{\natexlab{b}}, \apjl, 707, L142, \dodoi{10.1088/0004-637X/707/2/L142}

\bibitem[{{Ant{\'o}n} {et~al.}(2008){Ant{\'o}n}, {Browne}, \&
  {March{\~a}}}]{Anton08}
{Ant{\'o}n}, S., {Browne}, I.~W.~A., \& {March{\~a}}, M.~J. 2008, \aap, 490,
  583, \dodoi{10.1051/0004-6361:20078926}

\bibitem[{{Berton} {et~al.}(2016){Berton}, {Caccianiga}, {Foschini},
  {Peterson}, {Mathur}, {Terreran}, {Ciroi}, {Congiu}, {Cracco}, {Frezzato},
  {La Mura}, \& {Rafanelli}}]{Berton16c}
{Berton}, M., {Caccianiga}, A., {Foschini}, L., {et~al.} 2016, \aap, 591, A98,
  \dodoi{10.1051/0004-6361/201628171}

\bibitem[{{Berton} {et~al.}(2017){Berton}, {Foschini}, {Caccianiga}, {Ciroi},
  {Congiu}, {Cracco}, {Frezzato}, {La Mura}, \& {Rafanelli}}]{Berton17}
{Berton}, M., {Foschini}, L., {Caccianiga}, A., {et~al.} 2017, Frontiers in
  Astronomy and Space Sciences, 4, 8, \dodoi{10.3389/fspas.2017.00008}

\bibitem[{{Berton} {et~al.}(2018){Berton}, {Congiu}, {J{\"a}rvel{\"a}},
  {Antonucci}, {Kharb}, {Lister}, {Tarchi}, {Caccianiga}, {Chen}, {Foschini},
  {L{\"a}hteenm{\"a}ki}, {Richards}, {Ciroi}, {Cracco}, {Frezzato}, {La Mura},
  \& {Rafanelli}}]{Berton18a}
{Berton}, M., {Congiu}, E., {J{\"a}rvel{\"a}}, E., {et~al.} 2018, \aap, 614,
  A87, \dodoi{10.1051/0004-6361/201832612}

\bibitem[{{Boller} {et~al.}(1996){Boller}, {Brandt}, \& {Fink}}]{Boller96}
{Boller}, T., {Brandt}, W.~N., \& {Fink}, H. 1996, \aap, 305, 53

\bibitem[{{Boller} {et~al.}(1992){Boller}, {Meurs}, {Brinkmann}, {Fink},
  {Zimmermann}, \& {Adorf}}]{Boller92}
{Boller}, T., {Meurs}, E.~J.~A., {Brinkmann}, W., {et~al.} 1992, \aap, 261, 57

\bibitem[{{Boroson} \& {Green}(1992)}]{Boroson92}
{Boroson}, T.~A., \& {Green}, R.~F. 1992, \apjs, 80, 109,
  \dodoi{10.1086/191661}

\bibitem[{{Caccianiga} {et~al.}(2015){Caccianiga}, {Ant{\'o}n}, {Ballo},
  {Foschini}, {Maccacaro}, {Della Ceca}, {Severgnini}, {March{\~a}}, {Mateos},
  \& {Sani}}]{Caccianiga15}
{Caccianiga}, A., {Ant{\'o}n}, S., {Ballo}, L., {et~al.} 2015, \mnras, 451,
  1795, \dodoi{10.1093/mnras/stv939}

\bibitem[{{Chen} {et~al.}(2018){Chen}, {Berton}, {La Mura}, {Congiu}, {Cracco},
  {Foschini}, {Fan}, {Ciroi}, {Rafanelli}, \& {Bastieri}}]{Chen18}
{Chen}, S., {Berton}, M., {La Mura}, G., {et~al.} 2018, \aap, 615, A167,
  \dodoi{10.1051/0004-6361/201832678}

\bibitem[{{Chiaberge} {et~al.}(2015){Chiaberge}, {Gilli}, {Lotz}, \&
  {Norman}}]{Chiaberge15}
{Chiaberge}, M., {Gilli}, R., {Lotz}, J.~M., \& {Norman}, C. 2015, \apj, 806,
  147, \dodoi{10.1088/0004-637X/806/2/147}

\bibitem[{{Chiaberge} \& {Marconi}(2011)}]{Chiaberge11}
{Chiaberge}, M., \& {Marconi}, A. 2011, \mnras, 416, 917,
  \dodoi{10.1111/j.1365-2966.2011.19079.x}

\bibitem[{{Chiaberge} {et~al.}(2017){Chiaberge}, {Ely}, {Meyer},
  {Georganopoulos}, {Marinucci}, {Bianchi}, {Tremblay}, {Hilbert}, {Kotyla},
  {Capetti}, {Baum}, {Macchetto}, {Miley}, {O'Dea}, {Perlman}, {Sparks}, \&
  {Norman}}]{Chiaberge17}
{Chiaberge}, M., {Ely}, J.~C., {Meyer}, E.~T., {et~al.} 2017, \aap, 600, A57,
  \dodoi{10.1051/0004-6361/201629522}

\bibitem[{{Condon} {et~al.}(2002){Condon}, {Cotton}, \& {Broderick}}]{Condon02}
{Condon}, J.~J., {Cotton}, W.~D., \& {Broderick}, J.~J. 2002, \aj, 124, 675,
  \dodoi{10.1086/341650}

\bibitem[{{D'Ammando} {et~al.}(2018){D'Ammando}, {Acosta-Pulido}, {Capetti},
  {Baldi}, {Orienti}, {Raiteri}, \& {Ramos Almeida}}]{Dammando18}
{D'Ammando}, F., {Acosta-Pulido}, J.~A., {Capetti}, A., {et~al.} 2018, \mnras,
  478, L66, \dodoi{10.1093/mnrasl/sly072}

\bibitem[{{D'Ammando} {et~al.}(2017){D'Ammando}, {Acosta-Pulido}, {Capetti},
  {Raiteri}, {Baldi}, {Orienti}, \& {Ramos Almeida}}]{Dammando17}
---. 2017, \mnras, 469, L11, \dodoi{10.1093/mnrasl/slx042}

\bibitem[{{Decarli} {et~al.}(2008){Decarli}, {Dotti}, {Fontana}, \&
  {Haardt}}]{Decarli08}
{Decarli}, R., {Dotti}, M., {Fontana}, M., \& {Haardt}, F. 2008, \mnras, 386,
  L15, \dodoi{10.1111/j.1745-3933.2008.00451.x}

\bibitem[{{Deo} {et~al.}(2006){Deo}, {Crenshaw}, \& {Kraemer}}]{Deo06}
{Deo}, R.~P., {Crenshaw}, D.~M., \& {Kraemer}, S.~B. 2006, \aj, 132, 321,
  \dodoi{10.1086/504894}

\bibitem[{{Elizalde} \& {Steiner}(1994)}]{Elizalde94}
{Elizalde}, F., \& {Steiner}, J.~E. 1994, \mnras, 268,
  \dodoi{10.1093/mnras/268.1.L47}

\bibitem[{{Ferrarese} \& {Merritt}(2000)}]{Ferrarese00}
{Ferrarese}, L., \& {Merritt}, D. 2000, \apjl, 539, L9, \dodoi{10.1086/312838}

\bibitem[{{Fischer} {et~al.}(2006){Fischer}, {Iserlohe}, {Zuther}, {Bertram},
  {Straubmeier}, {Sch{\"o}del}, \& {Eckart}}]{Fischer06}
{Fischer}, S., {Iserlohe}, C., {Zuther}, J., {et~al.} 2006, \aap, 452, 827,
  \dodoi{10.1051/0004-6361:20053158}

\bibitem[{{Foschini}(2011)}]{Foschini11b}
{Foschini}, L. 2011, Research in Astronomy and Astrophysics, 11, 1266,
  \dodoi{10.1088/1674-4527/11/11/003}

\bibitem[{{Foschini} {et~al.}(2015){Foschini}, {Berton}, {Caccianiga}, {Ciroi},
  {Cracco}, {Peterson}, {Angelakis}, {Braito}, {Fuhrmann}, {Gallo}, {Grupe},
  {J{\"a}rvel{\"a}}, {Kaufmann}, {Komossa}, {Kovalev}, {L{\"a}hteenm{\"a}ki},
  {Lisakov}, {Lister}, {Mathur}, {Richards}, {Romano}, {Sievers},
  {Tagliaferri}, {Tammi}, {Tibolla}, {Tornikoski}, {Vercellone}, {La Mura},
  {Maraschi}, \& {Rafanelli}}]{Foschini15}
{Foschini}, L., {Berton}, M., {Caccianiga}, A., {et~al.} 2015, \aap, 575, A13,
  \dodoi{10.1051/0004-6361/201424972}

\bibitem[{{Goodrich}(1989)}]{Goodrich89}
{Goodrich}, R.~W. 1989, \apj, 342, 224, \dodoi{10.1086/167586}

\bibitem[{{Halpern} \& {Moran}(1998)}]{Halpern98}
{Halpern}, J.~P., \& {Moran}, E.~C. 1998, \apj, 494, 194,
  \dodoi{10.1086/305202}

\bibitem[{{Heinz} \& {Sunyaev}(2003)}]{Heinz03}
{Heinz}, S., \& {Sunyaev}, R.~A. 2003, \mnras, 343, L59,
  \dodoi{10.1046/j.1365-8711.2003.06918.x}

\bibitem[{{J{\"a}rvel{\"a}} {et~al.}(2018){J{\"a}rvel{\"a}},
  {L{\"a}hteenm{\"a}ki}, \& {Berton}}]{Jarvela18}
{J{\"a}rvel{\"a}}, E., {L{\"a}hteenm{\"a}ki}, A., \& {Berton}, M. 2018, ArXiv
  e-prints.
\newblock \doarXiv{1807.02970}

\bibitem[{{J{\"a}rvel{\"a}} {et~al.}(2017{\natexlab{a}}){J{\"a}rvel{\"a}},
  {L{\"a}hteenm{\"a}ki}, \& {Lietzen}}]{Jarvela17a}
{J{\"a}rvel{\"a}}, E., {L{\"a}hteenm{\"a}ki}, A., \& {Lietzen}, H.
  2017{\natexlab{a}}, Frontiers in Astronomy and Space Sciences, 4, 54,
  \dodoi{10.3389/fspas.2017.00054}

\bibitem[{{J{\"a}rvel{\"a}} {et~al.}(2017{\natexlab{b}}){J{\"a}rvel{\"a}},
  {L{\"a}hteenm{\"a}ki}, {Lietzen}, {Poudel}, {Hein{\"a}m{\"a}ki}, \&
  {Einasto}}]{Jarvela17}
{J{\"a}rvel{\"a}}, E., {L{\"a}hteenm{\"a}ki}, A., {Lietzen}, H., {et~al.}
  2017{\natexlab{b}}, \aap, 606, A9, \dodoi{10.1051/0004-6361/201731318}

\bibitem[{{Kawakatu} {et~al.}(2007){Kawakatu}, {Imanishi}, \&
  {Nagao}}]{Kawakatu07}
{Kawakatu}, N., {Imanishi}, M., \& {Nagao}, T. 2007, \apj, 661, 660,
  \dodoi{10.1086/516563}

\bibitem[{{Kellermann} {et~al.}(1989){Kellermann}, {Sramek}, {Schmidt},
  {Shaffer}, \& {Green}}]{Kellermann89}
{Kellermann}, K.~I., {Sramek}, R., {Schmidt}, M., {Shaffer}, D.~B., \& {Green},
  R. 1989, \aj, 98, 1195, \dodoi{10.1086/115207}

\bibitem[{{Komatsu} {et~al.}(2011){Komatsu}, {Smith}, {Dunkley}, {Bennett},
  {Gold}, {Hinshaw}, {Jarosik}, {Larson}, {Nolta}, {Page}, {Spergel},
  {Halpern}, {Hill}, {Kogut}, {Limon}, {Meyer}, {Odegard}, {Tucker}, {Weiland},
  {Wollack}, \& {Wright}}]{Komatsu11}
{Komatsu}, E., {Smith}, K.~M., {Dunkley}, J., {et~al.} 2011, \apjs, 192, 18,
  \dodoi{10.1088/0067-0049/192/2/18}

\bibitem[{{Komossa}(2018)}]{Komossa18}
{Komossa}, S. 2018, ArXiv e-prints.
\newblock \doarXiv{1807.03666}

\bibitem[{{Komossa} {et~al.}(2006){Komossa}, {Voges}, {Xu}, {Mathur}, {Adorf},
  {Lemson}, {Duschl}, \& {Grupe}}]{Komossa06}
{Komossa}, S., {Voges}, W., {Xu}, D., {et~al.} 2006, \aj, 132, 531,
  \dodoi{10.1086/505043}

\bibitem[{{Kormendy} \& {Ho}(2013)}]{Kormendy13}
{Kormendy}, J., \& {Ho}, L.~C. 2013, \araa, 51, 511,
  \dodoi{10.1146/annurev-astro-082708-101811}

\bibitem[{{Kotilainen} {et~al.}(2016){Kotilainen}, {Le{\'o}n-Tavares},
  {Olgu{\'{\i}}n-Iglesias}, {Baes}, {An{\'o}rve}, {Chavushyan}, \&
  {Carrasco}}]{Kotilainen16}
{Kotilainen}, J.~K., {Le{\'o}n-Tavares}, J., {Olgu{\'{\i}}n-Iglesias}, A.,
  {et~al.} 2016, \apj, 832, 157, \dodoi{10.3847/0004-637X/832/2/157}

\bibitem[{{Krongold} {et~al.}(2001){Krongold}, {Dultzin-Hacyan}, \&
  {Marziani}}]{Krongold01}
{Krongold}, Y., {Dultzin-Hacyan}, D., \& {Marziani}, P. 2001, \aj, 121, 702,
  \dodoi{10.1086/318768}

\bibitem[{{L{\"a}hteenm{\"a}ki} {et~al.}(2018){L{\"a}hteenm{\"a}ki},
  {J{\"a}rvel{\"a}}, {Ramakrishnan}, {Tornikoski}, {Tammi}, {Vera}, \&
  {Chamani}}]{Lahteenmaki18}
{L{\"a}hteenm{\"a}ki}, A., {J{\"a}rvel{\"a}}, E., {Ramakrishnan}, V., {et~al.}
  2018, \aap, 614, L1, \dodoi{10.1051/0004-6361/201833378}

\bibitem[{{Laor}(2000)}]{Laor00}
{Laor}, A. 2000, \apjl, 543, L111, \dodoi{10.1086/317280}

\bibitem[{{Leipski} {et~al.}(2007){Leipski}, {Haas}, {Siebenmorgen},
  {Meusinger}, {Albrecht}, {Cesarsky}, {Chini}, {Cutri}, {Drass}, {Huchra},
  {Ott}, \& {Wilkes}}]{Leipski07}
{Leipski}, C., {Haas}, M., {Siebenmorgen}, R., {et~al.} 2007, \aap, 473, 121,
  \dodoi{10.1051/0004-6361:20066323}

\bibitem[{{Le{\'o}n Tavares} {et~al.}(2014){Le{\'o}n Tavares}, {Kotilainen},
  {Chavushyan}, {A{\~n}orve}, {Puerari}, {Cruz-Gonz{\'a}lez},
  {Pati{\~n}o-Alvarez}, {Ant{\'o}n}, {Carrami{\~n}ana}, {Carrasco}, {Guichard},
  {Karhunen}, {Olgu{\'{\i}}n-Iglesias}, {Sanghvi}, \& {Valdes}}]{Leontavares14}
{Le{\'o}n Tavares}, J., {Kotilainen}, J., {Chavushyan}, V., {et~al.} 2014,
  \apj, 795, 58, \dodoi{10.1088/0004-637X/795/1/58}

\bibitem[{{Mao} {et~al.}(2015){Mao}, {Owen}, {Duffin}, {Keel}, {Lacy},
  {Momjian}, {Morrison}, {Mroczkowski}, {Neff}, {Norris}, {Schmitt}, {Toy}, \&
  {Veilleux}}]{Mao15}
{Mao}, M.~Y., {Owen}, F., {Duffin}, R., {et~al.} 2015, \mnras, 446, 4176,
  \dodoi{10.1093/mnras/stu2302}

\bibitem[{{Masci} {et~al.}(2010){Masci}, {Cutri}, {Francis}, {Nelson},
  {Huchra}, {Heath Jones}, {Colless}, \& {Saunders}}]{Masci10}
{Masci}, F.~J., {Cutri}, R.~M., {Francis}, P.~J., {et~al.} 2010, \pasa, 27,
  302, \dodoi{10.1071/AS10001}

\bibitem[{{Mathur}(2000)}]{Mathur00}
{Mathur}, S. 2000, \mnras, 314, L17, \dodoi{10.1046/j.1365-8711.2000.03530.x}

\bibitem[{{Mathur} {et~al.}(2012){Mathur}, {Fields}, {Peterson}, \&
  {Grupe}}]{Mathur12}
{Mathur}, S., {Fields}, D., {Peterson}, B.~M., \& {Grupe}, D. 2012, \apj, 754,
  146, \dodoi{10.1088/0004-637X/754/2/146}

\bibitem[{{Mathur} \& {Grupe}(2005{\natexlab{a}})}]{Mathur05a}
{Mathur}, S., \& {Grupe}, D. 2005{\natexlab{a}}, \aap, 432, 463,
  \dodoi{10.1051/0004-6361:20041717}

\bibitem[{{Mathur} \& {Grupe}(2005{\natexlab{b}})}]{Mathur05b}
---. 2005{\natexlab{b}}, \apj, 633, 688, \dodoi{10.1086/491613}

\bibitem[{{Mathur} {et~al.}(2001){Mathur}, {Kuraszkiewicz}, \&
  {Czerny}}]{Mathur01}
{Mathur}, S., {Kuraszkiewicz}, J., \& {Czerny}, B. 2001, \na, 6, 321,
  \dodoi{10.1016/S1384-1076(01)00058-6}

\bibitem[{{Monet} {et~al.}(2003){Monet}, {Levine}, {Canzian}, {Ables}, {Bird},
  {Dahn}, {Guetter}, {Harris}, {Henden}, {Leggett}, {Levison}, {Luginbuhl},
  {Martini}, {Monet}, {Munn}, {Pier}, {Rhodes}, {Riepe}, {Sell}, {Stone},
  {Vrba}, {Walker}, {Westerhout}, {Brucato}, {Reid}, {Schoening}, {Hartley},
  {Read}, \& {Tritton}}]{Monet03}
{Monet}, D.~G., {Levine}, S.~E., {Canzian}, B., {et~al.} 2003, \aj, 125, 984,
  \dodoi{10.1086/345888}

\bibitem[{{Olgu{\'{\i}}n-Iglesias} {et~al.}(2017){Olgu{\'{\i}}n-Iglesias},
  {Kotilainen}, {Le{\'o}n Tavares}, {Chavushyan}, \&
  {A{\~n}orve}}]{Olguiniglesias17}
{Olgu{\'{\i}}n-Iglesias}, A., {Kotilainen}, J.~K., {Le{\'o}n Tavares}, J.,
  {Chavushyan}, V., \& {A{\~n}orve}, C. 2017, \mnras, 467, 3712,
  \dodoi{10.1093/mnras/stx022}

\bibitem[{{Orban de Xivry} {et~al.}(2011){Orban de Xivry}, {Davies},
  {Schartmann}, {Komossa}, {Marconi}, {Hicks}, {Engel}, \&
  {Tacconi}}]{Orbandexivry11}
{Orban de Xivry}, G., {Davies}, R., {Schartmann}, M., {et~al.} 2011, \mnras,
  417, 2721, \dodoi{10.1111/j.1365-2966.2011.19439.x}

\bibitem[{{Osterbrock} \& {Pogge}(1987)}]{Osterbrock87}
{Osterbrock}, D.~E., \& {Pogge}, R.~W. 1987, \apj, 323, 108,
  \dodoi{10.1086/165810}

\bibitem[{{Padovani}(2017)}]{Padovani17}
{Padovani}, P. 2017, Nature Astronomy, 1, 0194, \dodoi{10.1038/s41550-017-0194}

\bibitem[{{Peng} {et~al.}(2002){Peng}, {Ho}, {Impey}, \& {Rix}}]{Peng02}
{Peng}, C.~Y., {Ho}, L.~C., {Impey}, C.~D., \& {Rix}, H.-W. 2002, \aj, 124,
  266, \dodoi{10.1086/340952}

\bibitem[{{Peng} {et~al.}(2010){Peng}, {Ho}, {Impey}, \& {Rix}}]{Peng10}
---. 2010, \aj, 139, 2097, \dodoi{10.1088/0004-6256/139/6/2097}

\bibitem[{{Peterson}(2011)}]{Peterson11}
{Peterson}, B.~M. 2011, in Narrow-Line Seyfert 1 Galaxies and their Place in
  the Universe, 32

\bibitem[{{Popovi{\'c}} {et~al.}(2004){Popovi{\'c}}, {Mediavilla}, {Bon}, \&
  {Ili{\'c}}}]{Popovic04}
{Popovi{\'c}}, L.~{\v C}., {Mediavilla}, E., {Bon}, E., \& {Ili{\'c}}, D. 2004,
  \aap, 423, 909, \dodoi{10.1051/0004-6361:20034431}

\bibitem[{{Porcel} {et~al.}(1998){Porcel}, {Garzon}, {Jimenez-Vicente}, \&
  {Battaner}}]{Porcel98}
{Porcel}, C., {Garzon}, F., {Jimenez-Vicente}, J., \& {Battaner}, E. 1998,
  \aap, 330, 136

\bibitem[{{Salucci} {et~al.}(2000){Salucci}, {Ratnam}, {Monaco}, \&
  {Danese}}]{Salucci00}
{Salucci}, P., {Ratnam}, C., {Monaco}, P., \& {Danese}, L. 2000, \mnras, 317,
  488, \dodoi{10.1046/j.1365-8711.2000.03622.x}

\bibitem[{{Sanders} {et~al.}(1988){Sanders}, {Soifer}, {Elias}, {Madore},
  {Matthews}, {Neugebauer}, \& {Scoville}}]{Sanders88}
{Sanders}, D.~B., {Soifer}, B.~T., {Elias}, J.~H., {et~al.} 1988, \apj, 325,
  74, \dodoi{10.1086/165983}

\bibitem[{{Sani} {et~al.}(2010){Sani}, {Lutz}, {Risaliti}, {Netzer}, {Gallo},
  {Trakhtenbrot}, {Sturm}, \& {Boller}}]{Sani10}
{Sani}, E., {Lutz}, D., {Risaliti}, G., {et~al.} 2010, \mnras, 403, 1246,
  \dodoi{10.1111/j.1365-2966.2009.16217.x}

\bibitem[{{Schmidt} \& {Green}(1983)}]{Schmidt83}
{Schmidt}, M., \& {Green}, R.~F. 1983, \apj, 269, 352, \dodoi{10.1086/161048}

\bibitem[{{Shen} \& {Ho}(2014)}]{Shen14}
{Shen}, Y., \& {Ho}, L.~C. 2014, \nat, 513, 210, \dodoi{10.1038/nature13712}

\bibitem[{{Sulentic} {et~al.}(2000){Sulentic}, {Zwitter}, {Marziani}, \&
  {Dultzin-Hacyan}}]{Sulentic00}
{Sulentic}, J.~W., {Zwitter}, T., {Marziani}, P., \& {Dultzin-Hacyan}, D. 2000,
  \apjl, 536, L5, \dodoi{10.1086/312717}

\bibitem[{{Wang} \& {Zhang}(2007)}]{Wang07}
{Wang}, J.-M., \& {Zhang}, E.-P. 2007, \apj, 660, 1072, \dodoi{10.1086/513685}

\bibitem[{{Williams} {et~al.}(2004){Williams}, {Mathur}, \&
  {Pogge}}]{Williams04}
{Williams}, R.~J., {Mathur}, S., \& {Pogge}, R.~W. 2004, \apj, 610, 737,
  \dodoi{10.1086/421768}

\bibitem[{{Xu} {et~al.}(2012){Xu}, {Komossa}, {Zhou}, {Lu}, {Li}, {Grupe},
  {Wang}, \& {Yuan}}]{Xu12}
{Xu}, D., {Komossa}, S., {Zhou}, H., {et~al.} 2012, \aj, 143, 83,
  \dodoi{10.1088/0004-6256/143/4/83}

\bibitem[{{Xu} {et~al.}(1999){Xu}, {Wei}, \& {Hu}}]{Xu99}
{Xu}, D.~W., {Wei}, J.~Y., \& {Hu}, J.~Y. 1999, \apj, 517, 622,
  \dodoi{10.1086/307239}

\end{thebibliography}

\end{document}